\documentclass[aps, prb, showpacs, reprint, superscriptaddress]{revtex4-1}

\usepackage{graphicx}
\usepackage{amssymb,amsmath}
\usepackage{bm}
\usepackage{float}
\usepackage{xcolor}
\usepackage{mathrsfs}
\usepackage{extarrows}
\usepackage{bbm}
\definecolor{midnightblue}{cmyk}{1,1,0,0.1}
\definecolor{forestgreen}{cmyk}{0.75,0,1,0.5}

\usepackage{lineno}

\usepackage{hyperref}
\hypersetup{
    bookmarks=true,         
    unicode=false,          
    pdftoolbar=true,        
    pdfmenubar=true,        
    pdffitwindow=false,     
    pdfstartview={FitH},    
    pdftitle={My title},    
    pdfauthor={Author},     
    pdfsubject={Subject},   
    pdfcreator={Creator},   
    pdfproducer={Producer}, 
    pdfkeywords={keyword1} {key2} {key3}, 
    pdfnewwindow=true,      
    colorlinks=true,       
    linkcolor=blue,          
    citecolor=blue,        
    filecolor=blue,      
    urlcolor=blue,          
}

\def\sint{\ifmmode{- \!\!\!\!\!\! \int}
    \else{\hbox{$- \!\!\!\! \int \ $}}\fi}

\begin{document}
\preprint{Physical Review B}
\title{$(111)$ Surface States of SnTe}

\author{Yin Shi}
\affiliation{International Center for Quantum Materials, Peking University, Beijing 100871, China.}
\affiliation{Collaborative Innovation Center of Quantum Matter, Beijing, China}

\author{Meng Wu}
\affiliation{International Center for Quantum Materials, Peking University, Beijing 100871, China.}
\affiliation{Collaborative Innovation Center of Quantum Matter, Beijing, China}

\author{Fan Zhang}
\email{zhang@utdallas.edu}
\affiliation{
 Department of Physics, The University of Texas at Dallas, Richardson, Texas 75080, USA
}
\author{Ji Feng}
\email{jfeng11@pku.edu.cn}
\affiliation{International Center for Quantum Materials, Peking University, Beijing 100871, China.}
\affiliation{Collaborative Innovation Center of Quantum Matter, Beijing, China}

\date{\today}

\begin{abstract}
The characterization and applications of topological insulators depend critically on their protected surface states,
which, however, can be obscured by the presence of trivial dangling bond states.
Our first principle calculations show that this is the case for the pristine $(111)$ surface of SnTe.
Yet, the predicted surface states unfold when the dangling bond states are passivated in proper chemisorption.
We further extract the anisotropic Fermi velocities, penetration lengths and anisotropic spin textures of the unfolded $\bar\Gamma$- and $\bar M$-surface states, which are consistent with the theory in \href{http://dx.doi.org/10.1103/PhysRevB.86.081303}{Phys. Rev. B 86, 081303 (R)}.
More importantly, this chemisorption scheme provides an external control of the relative energies of different Dirac nodes,
which is particularly desirable in multi-valley transport.
\end{abstract}

\maketitle
\section{Introduction}
\begin{figure}
\centering
\includegraphics[width=0.45\textwidth]{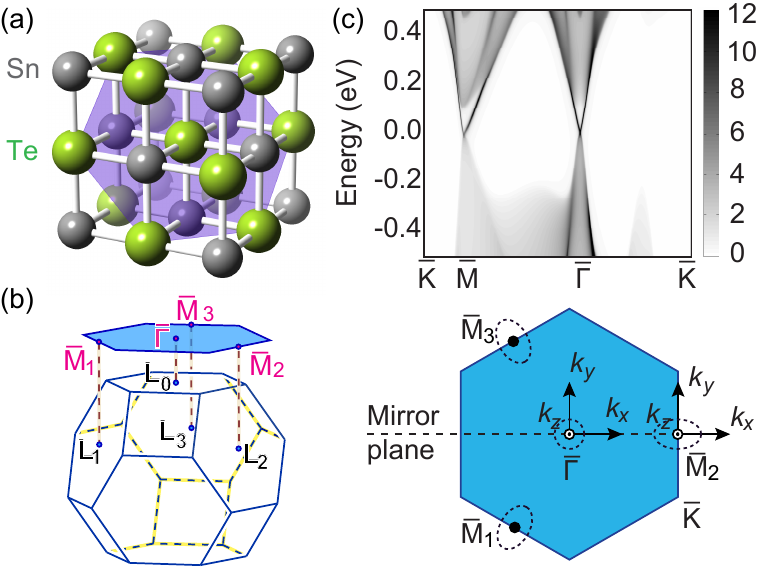}
\caption{(a) The crystal structure of bulk SnTe. The light-blue plane is a $(111)$ lattice plane with only Te atoms.
(b) The bulk BZ and the $(111)$ surface BZ.
(c) Band structure of SnTe semi-infinite $(111)$ slab calculated in the iterative Green's function method.
The gray value represents the momentum-resolved density of states. The surface BZ and its high symmetry points are shown in the lower panel, where the dotted lines represent the isoenergy contours around Dirac cones.}
\label{fig:structure}
\end{figure}

The advent of $Z_2$ topological insulators (TIs) protected by time-reversal symmetry~\citep{HasanKane10,Moore10,QiZhang10}
opened the door to the search for other topological states with various different symmetries.
Recently, tin telluride (SnTe) was predicted by Hsieh {\it et al.}~\cite{Hsieh12}
to be a representative three-dimensional TI protected by mirror symmetries.
Subsequently, the predicted $(001)$ surface states were observed in angle-resolved photoemission spectroscopy (ARPES)~\cite{Tanaka12,Dziawa12,Xu12}.
SnTe has the rock salt crystal structure, as shown in Fig.~\ref{fig:structure}(a), with a bulk energy gap about $0.3$~eV near the $L$ points.
Importantly, the band gap is inverted at the four inequivalent $L$ points.
Although the strong and weak $Z_2$ indices are all zero in the SnTe case, a set of nontrivial mirror Chern numbers exists~\cite{Hsieh12}
in the presence of $(110)$-like mirror symmetries. As a consequence~\cite{Hsieh12,Zhang13},
any surface respecting the mirror symmetry hosts even number of gapless Dirac surface states.
Considerable theoretical and experimental efforts have focused on the $(001)$ surface,
where four Dirac cones are observed near the surface Brillouin zone (BZ) boundaries~\cite{Hsieh12,Tanaka12,Dziawa12,Xu12}.

In contrast, the more exotic $(111)$ surface band structure~\cite{Safaei13,Zhang13} has been relatively unexplored.
The coexistence of multiple, symmetry-related and symmetry-unrelated, isotropic and anisotropic, surface Dirac cones on the $(111)$ surface
[Fig.~\ref{fig:structure}(c)]
in fact may lead to remarkable valley contrasting physics, e.g., tunable Chern insulators with surface magnetization~\cite{Zhang13}
and designer topological insulators in superlattices~\cite{Li13}.
More recently, Y. Tanaka {\it et al.} have experimentally explored the SnTe $(111)$ surface states~\cite{Tanaka13}.
While confirming the existence of Dirac cones centered at $\bar{\Gamma}$ and $\bar{M}$, as depicted in Fig.~\ref{fig:structure}(c),
they found that the relative energy position of the Dirac points at $\bar{\Gamma}$ and $\bar{M}$ is reversed compared with some TB results~\cite{Safaei13}.

Thus, treatment beyond independent-electron approximations is urgent and necessary,
especially given the aforementioned discrepancy in energetic ordering of surface Dirac points.
A recent work~\cite{Wang14} used the density-functional theory (DFT) method to model the reconstruction of SnTe $(111)$ surface, in which three stable surface phases were established.
Distinct from this effort, here we will focus on effects of the intrinsic and extrinsic surface chemistry on SnTe $(111)$ surface states.
Particularly, we are motivated to study the resulting properties of the surface states, especially the tunable energetic ordering of different surface Dirac points, which has been observed in a recent experiment~\cite{Tanaka13} and likely to be important for future applications in electronics.
In order to elucidate them, we present a systematic study of the electronic structure of the SnTe $(111)$ surface based on DFT calculations,
with close comparisons with low-energy continuum theories~\cite{Zhang13,Zhang12}, tight-binding calculations~\cite{Safaei13}, and ARPES experiments~\cite{Tanaka13}.
We first show that the pristine Sn- and Te-terminated $(111)$ surface states comprise of multiple bands cluttering up the bulk gap.
We then demonstrate in detail that surface chemistry can play a key role in tailoring the topological surface states.
In particular, the surface chemisorption can retrieve the protected surface Dirac cones at $\bar{\Gamma}$ and $\bar{M}_i$ ($i=1,2,3$)~\cite{Wang14},
as anticipated based on the bulk electronic structure~\cite{Zhang13,Safaei13}, by repelling the trivial dangling bond states away from the bulk energy gap.
We further extract the anisotropic Fermi velocities, penetration lengths, and anisotropic spin textures of the unfolded $\bar\Gamma$ and $\bar M$ surface Dirac states, which are consistent with the theory in Ref.~\onlinecite{Zhang12}.
More importantly, with different adatoms we numerically and theoretically show that the relative energy position between the $\bar{\Gamma}$ and $\bar{M}$ Dirac points can be tuned~\cite{Zhang12} via the surface chemistry, which is novel.

\section{Beyond Dirac surface states}
We start from constructing a bulk TB model using Wannier representation of Kohn-Sham Bloch states,
and then employ the iterative Green's function~\cite{Lopez84,Lopez85} to compute the surface states.
The surface band structure is revealed by the imaginary part of the surface Green's function,
which can be viewed as a momentum-resolved surface density of states (DOS).
Fig.~\ref{fig:structure}(c) shows the gapless Dirac surface bands at $\bar{\Gamma}$ and $\bar{M}$ points,
which are consistent with the predictions based on a TB model~\cite{Safaei13} and a continuum model~\cite{Zhang13}.
We note that the iterative Green's function calculations as well as previous TB and continuum theories~\cite{Safaei13,Zhang13}
are likely not adequate to describe the bonding of real surfaces,
as the surface chemistry, namely adsorption and reconstruction, is absent in these theories.
Evidently, our following DFT method constitutes the advance of incorporating the surface chemistry.

\begin{figure}
\centering
\includegraphics[width=0.4\textwidth]{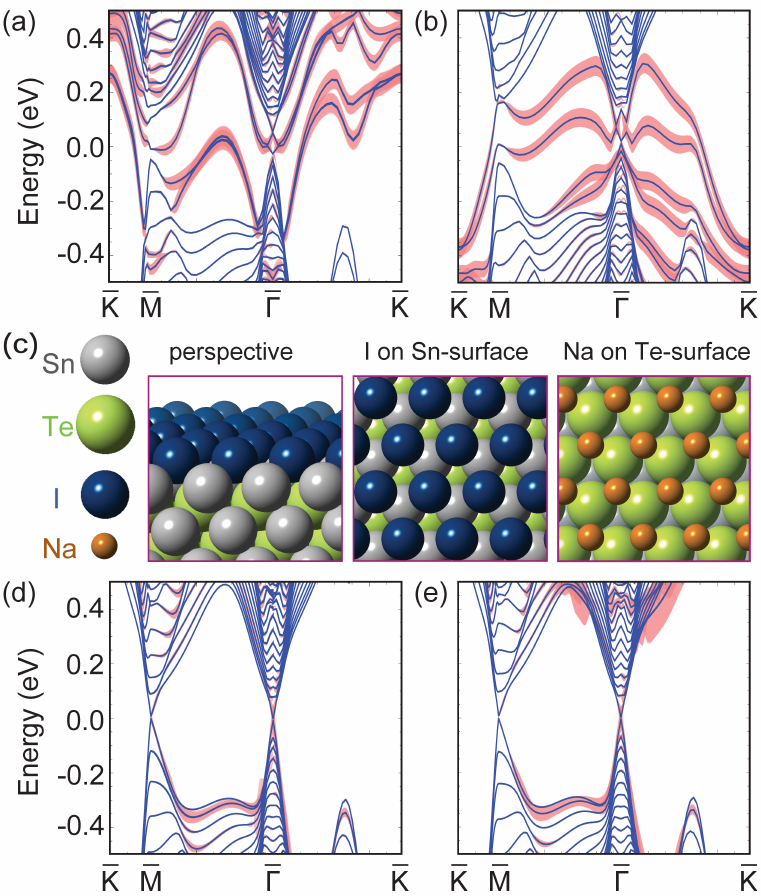}
\caption{DFT band structures of SnTe $(111)$ slabs. (a) Pristine Sn-terminated surface. (b) Pristine Te-terminated surface.
(c) Adsorption geometry for iodine on the Sn-terminated surface and that for sodium on the Te-terminated surface.
Electronic structures of (d) I-Sn surface, and (e) Na-Te surface.
In all panels for band structures, the vertical width of a pink fat band shows the extent of its localization near the surface.
The fat band width in (d) and (e) has been magnified by three times, compared with that in (a) and (b).}
\label{fig:chemisorption}
\end{figure}

We perform DFT calculations with the generalized-gradient approximation and the Perdew-Burke-Ernzerhof exchange-correlation functional,
using the projector-augmented wave potentials~\cite{Blochl94,Perdew96,Kresse93,Kresse94,Kresse96,Kresse96Oct}.
A plane-wave kinetic energy cutoff of $229$~eV is used in all calculations and the spin-orbit coupling is included non-self consistently.
After confirming the agreement of our bulk SnTe band structure with a previous report~\cite{littlewood10},
we systematically study the electronic structure of the $(111)$ surface in slab geometry.
To model real surface conditions in our DFT calculations of $(111)$ slab, we have optimized the positions of
the atoms from the first four atomic layers (as well as the adatoms in a later case) while fixing other interior atoms.
Fig.~\ref{fig:chemisorption}(a) and (b) respectively show the band structures of
pristine Sn- and Te-terminated $(111)$ slabs with $79$ atomic layers.
The width of superimposed fat bands indicates the extent of their localization near the surface.
We see that both of the two pristine surfaces have $8$ surface bands cluttering up the bulk gap,
reflecting the dangling bond states of the unsaturated clean surface.
Evidently, the surface band structure from our DFT calculations is in sharp contrast to
the result from the Green's function method, in which the dangling bond states cannot be captured.
As SnTe only has an even number of Dirac surface states on a mirror symmetric surface,
the presence of multiple dangling bond states may break the translational and mirror symmetries,
and thus couple and gap the Dirac surface states.
This will make the interpretation of ARPES, and particularly, transport results difficult.

\section{Passivation of dangling bond states}
It is highly desirable, therefore, to eliminate the non-topological dangling bond states
from the bulk gap to reveal the protected surface states.
It is natural to suggest that adsorption of chemical species will saturate the dangling bonds and thus help the topological surface states unfold.
The criterion for choosing the proper adsorption species can be established
with simple electron counting based on the octet rule of covalence.
Chemically, we can think of a formal valence $+2$ for Sn and $-2$ for Te.
It follows that a Sn-terminated surface has $1$ electron per Sn to donate or share covalently,
and that a monolayer of halogen will be suitable for the surface state passivation.
Similarly, a Te-terminated surface will grab an extra electron per Te from the suitable adatoms, e.g., the hydrogen atom~\cite{Wang14} or alkali metals.
In this section, we choose iodine (sodium) on the Sn (Te)-terminated surface to demonstrate this passivation, since I$^{-1}$ (Na$^{+1}$) and Sn$^{+2}$ (Te$^{-2}$) are close in size.
(Ref~\onlinecite{Wang14} used hydrogen adatom atop surface Te to execute the passivation;
however, since H$^{+1}$ ion's size is way smaller than that of Te$^{-2}$, hydrogen atoms may be drawn into the outmost layer of Te atoms, which may not fully saturate the surface dangling bonds.)

To find out the optimal adsorption site, we compare the adsorption energies
$$E_{\text{ad}}=E_{\text{SnTe-adatom}}-E_{\text{SnTe}}-E_{\text{adatom}}$$
of possible adsorption sites on the Sn or Te triangular lattice, allowing full relaxation of the adatom positions.
The number of SnTe atomic layers included in the calculations is the same as that in Fig.~\ref{fig:chemisorption}(a) and (b).
We find that the triangular center [Fig.~\ref{fig:chemisorption}(c)] is most stable
for both iodine atoms on the Sn-terminated surface and sodium atoms on the Te-terminated surface,
with adsorption energy $-22$ and $-12$~meV/atom, respectively.
It is quite gratifying to find that the optimized adsorption geometry maintains the {\it three} mirror symmetries of the $(111)$ surface,
which is the key to protect gapless Dirac surface states.
As shown in Fig.~\ref{fig:chemisorption}(d) and (e), after passivation,
we can clearly identify four Dirac cones at $\bar{M}_i$ ($i=1,2,3$) and $\bar{\Gamma}$ points on both Sn- and Te-terminated surfaces.

\begin{figure}[t!]
\centering
\includegraphics[width=0.45\textwidth]{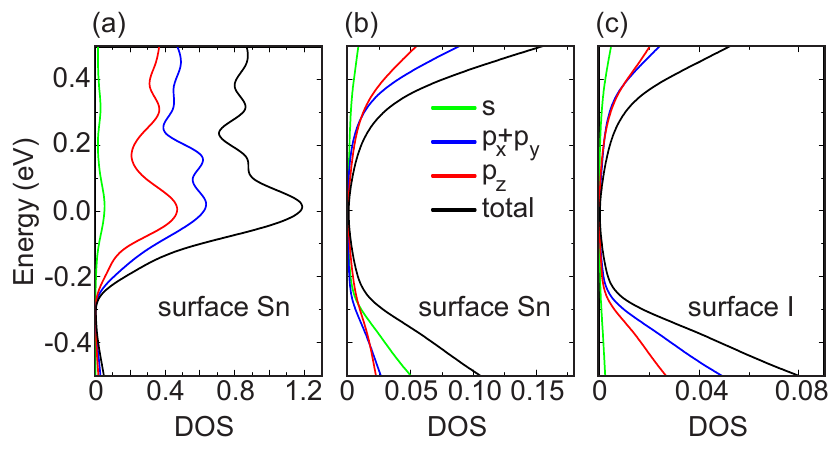}
\caption{DOS projected onto the $s$, $p_x+p_y$, $p_z$ orbits of surface atoms. The DOS of $p_x$ and $p_y$ are the same.
(a) Projected DOS on the Sn atoms at the bottom surface of pristine SnTe $(111)$ slab with $79$ atomic layers.
(b) The same as (a) but the surface is decorated with iodine.
(c) Projected DOS on the bottom iodine adatoms of the decorated slab.}
\label{fig:DOS}
\end{figure}

In order to elucidate the mechanism of chemical extraction of protected Dirac surface states,
we compare the projected DOS of the pristine Sn-terminated surface and the halogenated surface in Fig.~\ref{fig:DOS}.
In the absence of chemisorption, the $p$-orbital states of surface Sn atoms are not completely paired,
forming bands near the Fermi energy, as shown in Fig.~\ref{fig:DOS}(a).
In contrast, on a halogenated surface the dangling bond states are repelled away from the bulk gap
by forming bonding and anti-bonding states with the adsorbed chemical species.
This can be seen in Fig.~\ref{fig:DOS}(b) and (c), in which the DOS $\sim E$ provides another evidence of the unfolding of Dirac surface states in the bulk gap.
Hence, the four eliminated surface bands correspond to the unpaired $p$-orbitals from the top and bottom surfaces
whereas the four remaining Dirac-cone-like bands are the consequence of the nontrivial bulk topological invariant.

\section{Fermi velocities}  \label{sec:FV}
With chemical passivation of the dangling bond states,
it becomes possible to examine the intrinsic properties of the topological surface states
by fitting our DFT results to results from the continuum models~\cite{Zhang12,Zhang13} and from the experiments~\cite{Tanaka12,Dziawa12,Xu12,Tanaka13}.
The surface Dirac cone at $\bar{\Gamma}$  point is isotropic whereas those at $\bar{M}$ points are anisotropic~\cite{Zhang12,Zhang13,Tanaka13}.
For the I-Sn surface, we obtain three different Fermi velocities, namely, the Fermi velocity of $\bar\Gamma$ Dirac cone $v_{\bar{\Gamma}}=3.04$, the Fermi velocity of $\bar M$ Dirac cone along $\overline{\bar M \bar K}$ $v_{\bar{M}\bar{K}}=2.90$, and the Fermi velocity of $\bar M$ Dirac cone along $\overline{\bar M \bar \Gamma}$ $v_{\bar{M}\bar{\Gamma}}=1.68$~eV$\cdot$\AA.

On the other hand, we can also obtain the Fermi velocities from our DFT calculation of the bulk valence band, $v_z=1.89$ and $v_y=2.90$~eV$\cdot$\AA,
which are defined in the bulk $\bm{k}\cdot\bm{p}$ Hamiltonian at each $L$ point~\cite{Hsieh12,Zhang12}
\begin{eqnarray}
\mathcal{H}_L &=& m\sigma_z + v_z k_z\sigma_y+v_y(k_y s_x - k_x s_y)\sigma_x\,. \label{eq:HL}
\end{eqnarray}
Here $m\approx 0.3$~eV is the bulk band gap at $L$ point, $\hat{k}_z=\overline{\Gamma L}$, and $\hat{k}_y$ is normal to the $k_x$-$k_z$ mirror.
$\bm{s}$ are the real spin Pauli matrices whereas $\sigma_z=\pm$ denote the Sn and Te $p$-orbital pseudospins.
According to a theory that is applicable to any crystal face of SnTe~\cite{Zhang12,Zhang13},
$v_{\bar{\Gamma}}=v_{\bar{M}\bar{K}}=v_y$ and
$v_{\bar{M}\bar{\Gamma}}=v_zv_y/\sqrt{(v_z\cos\theta_{\bar M})^2+(v_y\sin\theta_{\bar M})^2}$ with $\cos\theta_{\bar M}=1/3$,
which approximately hold in our DFT results.

\begin{figure}[t!]
\centering
\includegraphics[width=0.45\textwidth]{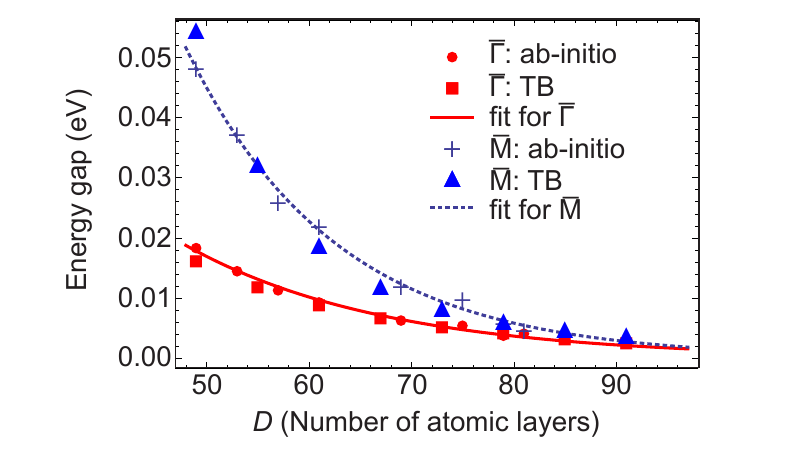}
\caption{Surface state energy gaps at $\bar{\Gamma}$ and $\bar{M}$, induced by the hybridization between the top and bottom surfaces,
as a function of the slab thickness. The solid and dashed lines are the exponential fitting with $E_g^0 e^{-D/2l_0}$.}
\label{fig:TG}
\end{figure}

The Dirac surface states can penetrate into the bulk. As a consequence,
For thin slabs the top and bottom surface states can hybridize and induce a surface band gap.
We compute the hybridized surface band gaps of a series of iodine-passivated Sn-terminated slabs with increasing number of atomic layers in Fig.~\ref{fig:TG} 
{\color{red}
(The gaps for the number of atomic layers exceeding 81 are only calculated within the TB model)}.
The hybridized gaps decay exponentially with increasing thickness.
The gap at $\bar{\Gamma}$ becomes negligible ($<1$~meV) when the thickness exceeds $106$ atomic layers.
This value is relatively thicker compared with the one for Bi$_2$Se$_3$ $(111)$ slabs,
in which $30$ atomic layers is sufficient to close the hybridized gap~\cite{Yazyev10}.
We note that the $\bar{\Gamma}$ and $\bar{M}$ surface states have different penetration lengths, i.e., $l_0 = 9.8$ and $7.3$ atomic layers, respectively.
This contrast in penetration length may have important implications in future valley engineering of the SnTe surface states.
$2l_0$ is also comparable to $\hbar v/{E_g^0}$, the result from a continuum model~\cite{Zhang12}.

\section{Spin Textures}  \label{sec:ST}
\begin{figure}[t!]
\centering
\includegraphics[width=0.45\textwidth]{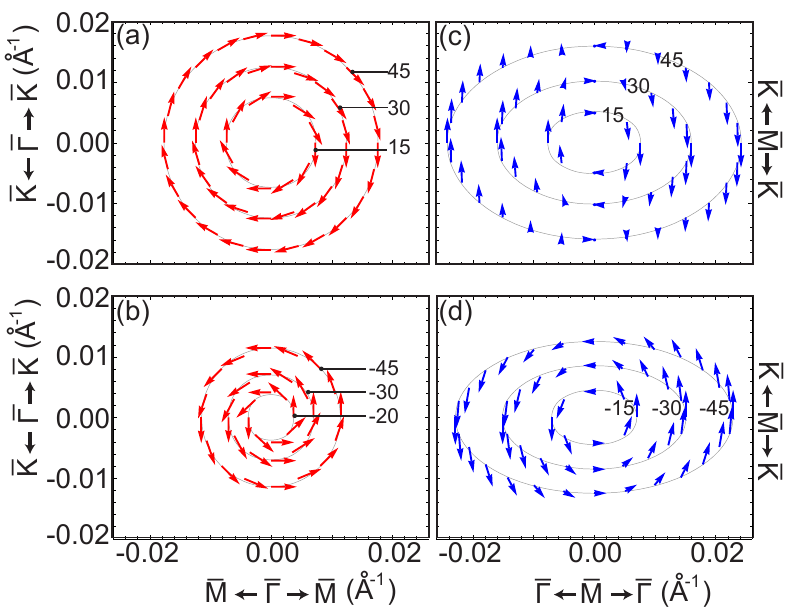}
\caption{The in-plane projection of the surface state spin textures.
(a) Conduction-band textures at $\bar{\Gamma}$; (b) Valence-band textures at $\bar{\Gamma}$;
(c) conduction-band textures at $\bar{M}$; (d) valence-band textures at $\bar{M}$.
The numbers denote the energies of constant energy contours in units of meV with the reference being the Fermi energy.}
\label{fig:spintextureabinio}
\end{figure}
Our DFT calculations also reveal the spin textures of $\bar\Gamma$ and $\bar M$ Dirac surface states,
as shown in Fig.~\ref{fig:spintextureabinio}.
Like the case in $\rm Bi_2Se_3$, both $\bar \Gamma$ and $\bar M$ surface states have
opposite pseudospin helicities (or winding numbers) for the conduction and valence bands~\cite{Zhang12}.
The isotropic surface state at $\bar\Gamma$ point is reminiscent of the cleavage surface state of $\rm Bi_2Se_3$,
whereas the anisotropic surface state at $\bar M$ point recalls the previous predictions~\cite{Zhang12,Safaei13}.
These striking features can be well understood by the following spin texture formula~\cite{Zhang12}
of a Dirac surface state at a general crystal face:
\begin{flalign}
\label{spin}
&\langle s_x,s_y,s_z\rangle\!\!=\!\!\frac{\pm v_z v_y k_y\cos\theta}{v_3\sqrt{v_x^2k_x^2+v_y^2k_y^2}},
\frac{\mp v_z v_y k_x}{v_3\sqrt{v_x^2k_x^2+v_y^2k_y^2}},0\,.&
\end{flalign}
where $+\,\,(-)$ denotes the conduction (valence) band.
Note that the $\bm s$ quantization axes are defined in Eq.~(\ref{eq:HL}) of the original bulk $L_0$ ($L_2$) point for $\bar\Gamma$ ($\bar{M}_2$) Dirac cone,
whereas the axes of the right hand side of Eq.~(\ref{spin}) are defined on the surface with
$v_3=\sqrt{(v_z\cos\theta)^2+(v_y\sin\theta)^2}$ and $v_x=v_z v_y/v_3$.
For the $\bar\Gamma$ surface state $\cos\theta=\cos\theta_{\bar \Gamma}=1$ whereas for the $\bar M$ surface state $\cos\theta=\cos\theta_{\bar M}=1/3$.
As shown in Fig.~\ref{fig:spintextureabinio}, the spin texture is almost in-plane for the $\bar\Gamma$ surface state,
whereas it generally has an out-of-plane component for the $\bar M$ surface state. Here the plane refers to the $(111)$ surface.
In the $\bar M$ Dirac cone, the spin is a unit and completely in-plane at $k_{\bar{M}\bar{K}}=0$,
whereas it is tilted completely out-of-plane and less than a unit at $k_{\bar{M}\bar{\Gamma}}=0$.

\section{Surface potentials}  \label{sec:SP}
Now we evaluate the influence of chemisorption on the surface state energies, in particular,
to reveal the possibility of tuning the energy difference between $\bar{\Gamma}$ and $\bar{M}$ Dirac nodes,
$\delta E= E_{\bar{\Gamma}}-E_{\bar{M}}$.
Fundamentally, there is no symmetry that relates the $\bar{\Gamma}$ and $\bar{M}$ surface states and their Dirac point energies are not required to be the same.
Previous TB calculations~\cite{Liu13, Safaei13} and our Green's function results in Fig.~\ref{fig:structure}(c) both give $\delta E>0$.
Although $\delta E=10$~meV is small in Fig.~\ref{fig:structure}(c),
it reflects~\cite{Zhang12} the bulk particle-hole symmetry breaking and its intrinsic anisotropy
in the directions parallel and perpendicular to $(111)$ surface.
The higher {\it intrinsic} Dirac point energy at $\bar \Gamma$ is consistent~\cite{Zhang12} with that
the Dirac cone at $\bar\Gamma$ exhibits stronger particle-hole asymmetry than the one at $\bar M$,
as already shown in Fig.~\ref{fig:spintextureabinio}.

In our DFT calculations, however, the I-passivated and the Br-passivated Sn-terminated surfaces have $\delta E= -8$ and $-20$~meV, respectively.
This is quite counterintuitive, as it seems that a uniform monolayer
does not distinguish between valleys at $\bar{\Gamma}$ and $\bar{M}$.
Importantly, our DFT result $\delta E <0$ on both chemically passivated surfaces
is very consistent with the recent experiment by Tanaka {\it et al}.~\cite{Tanaka13}.
This consistency not only suggests that the experimentally prepared surface is passivated
(the polar surface is likely to be passivated by residual gases in a very short time),
but also demonstrates the tunability of $\delta E$ via surface potentials,
which may be critical to valley engineering in chemical means.

Insights into the chemical tuning of $\bar{\Gamma}$ and $\bar{M}$ Dirac points can be gained by
analyzing the surface perturbations allowed by the essential symmetries~\cite{Zhang12}.
The SnTe $(111)$ surface exhibits $C_{3v}$ point-group symmetry as well as the time-reversal symmetry, which is also preserved by the adatoms.
The $C_{3}$ symmetry relates the three $\bar M$ Dirac points and requires them to have the same energy.
We can thus focus on $\bar{\Gamma}$ and $\bar{M}_2$ Dirac points on the mirror normal to $\hat{k}_y$, as shown in the lower panel of Fig.~\ref{fig:structure}(c).
The symmetry restriction immediately leads to only three types of surface potentials to leading order: 
$\sigma_0$, $\sigma_x$ and $\sigma_z$.
Note that only $\sigma_x$ potential can change $\delta E$~\cite{Zhang12}. 
Thus, we only focus on one type of surface potentials, $\eta\delta(z)2v_z\cdot\sigma_x$ with $\eta$ in units of the bulk gap $m$ and $\delta(z)$ implying localization at $(111)$ surface.
Note that here $\hat{z}$ is normal to $(111)$ surface, as shown in the lower panel of Fig.~\ref{fig:structure}(c).
This $\sigma_x$ potential represents the change in the hopping amplitude between the surface Sn and Te,
and it is naturally negative ($\eta<0$) when induced by the adatoms on Sn or Te layer.
Solving the eigenvalues of Hamiltonian $[\mathcal{H}_L+\eta\delta(z)2v_z\cdot\sigma_x]$ using the topological boundary condition~\cite{Zhang12}, we obtain the net energy shift between $\bar \Gamma$ and $\bar M$ Dirac nodes:
\begin{equation}
\delta E = 10~\text{meV}+\frac{4\eta(1+\eta^2)m}{(1+\eta^2)^2+4\eta^2}-\frac{4\eta'(1+\eta'^2)m}{(1+\eta'^2)^2+4\eta'^2}\,,
\end{equation}
where $\eta'=\eta\alpha v_z/v_3$ with $\alpha=v_z\cos\theta_{\bar M}/v_3$.
Since intrinsically $v_z<v_3$ and $\alpha<1$ the extrinsic Dirac point energy difference, $(\delta E-10~\text{meV})$, is always negative for a small surface perturbation ($|\eta|<1$).
This analysis is consistent with our DFT results, $-18$ ($-30$)~meV induced by the iodine (bromine) adatom layer.
Bromine induces a more negative $\delta E$ since it has a stronger electronegativity, compared with iodine.

\section{Discussions}

\begin{figure}[t!]
\centering
\includegraphics[width=0.45\textwidth]{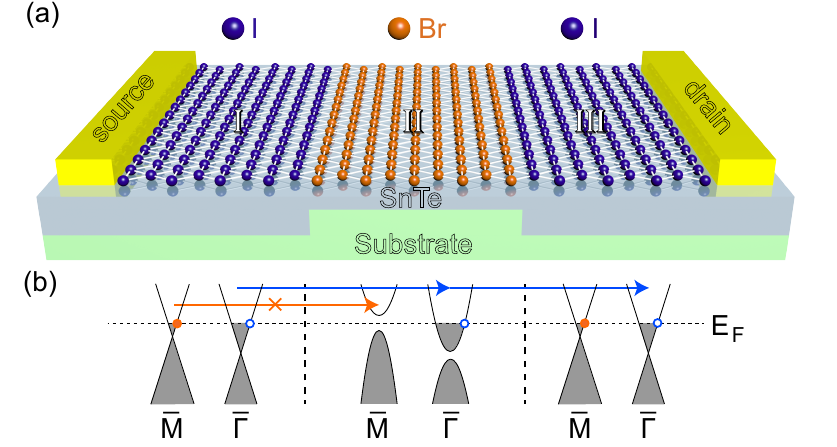}
\caption{(a) Schematic diagram of the $\bar{\Gamma}$ valley filter, where the top surface of the SnTe film is (111) surface. (b) Schematic plot of $\bar{\Gamma}$ valley selecting mechanism. The blue open (red filled) circles denote the electrons in $\bar{\Gamma}$ ($\bar{M}$) valley.}
\label{fig:device}
\end{figure}

The $(001)$ surface states of SnTe have been successfully probed by experiments~\cite{Xu12,Dziawa12,Tanaka12},
and our first-principles calculations concerning the pristine $(001)$ surface reveal no trivial surface states around the bulk gap.
In fact, the $(001)$ surface has a relatively high bonding saturation when compared with $(111)$ surface,
which moves the $p$-orbitals of surface atoms out of bulk energy gap.
This nature of the $(001)$ surface may be closely related to the successful probes of the four protected Dirac cones.
Hence we infer that the $(111)$ surface states are likely to be better probed with suitable surface chemisorption.
According to the $(111)$ surface free energy of pristine Sn- and Te-terminations~\cite{Wang14}, the latter is energetically preferable, which is likely to be the case in a recent transport examination experiment for SnTe $(111)$ surface prepared by molecular-beam epitaxy~\cite{Taskin14}.
However, we note that both terminations are indeed possible in the presence of surface passivation by other atoms.

More importantly, the chemisorption induces surface potential and can tune~\cite{Zhang12} the Dirac point energy difference
between the $\bar\Gamma$ and $\bar{M}$ surface states.
This energy difference may lead to a charge transfer and redistribution between the four valleys.
One may think of a momentum space p-n junction formed by one n-type $\bar\Gamma$ Dirac cone and three p-type $\bar{M}$ Dirac cones.
Also a $\bar{\Gamma}$ valley filter can be possibly designed, which is schematically shown in Fig. \ref{fig:device}.
The SnTe film at region II is fabricated to be thin so that hybridization gaps of surface bands at $\bar{\Gamma}$ and $\bar{M}$ result.
With the help of $\delta E$'s difference between iodine-adsorbed and bromine-adsorbed surface, proper doping can lead to a charge distribution as in Fig. \ref{fig:device}(b), where the $\bar\Gamma$ ($\bar M$) valley electrons can easily (hardly) transport from region I to region II, producing a $\bar\Gamma$ valley filter. Note that the bands in Fig. \ref{fig:device}(b) are doubly degenerate due to the assumed inversion symmetry for simplicity.

Moreover, the anisotropic Dirac cone may also lead to more intriguing surface plasmons than the case of graphene or $\rm Bi_2Se_3$,
providing an attractive alternative to noble-metal plasmons due to
their tighter confinement, anisotropic linear dispersions, and longer propagation distance.

The spin texture that we have identified for the $\bar{M}$ surface state
in fact represents a more general feature~\cite{Zhang12} for the Dirac surface state of a bulk material with $C_{3v}$ point-group symmetry,
compared with the texture of the $\bar\Gamma$ surface state or a similar one for $\rm Bi_2Se_3$.
The intrinsic anisotropy in the spin texture implies anomalous Zeeman coupling to spin~\cite{Zhang13Surface},
which may give rise to new phenomena in spintronics and valleytronics on the $(111)$ surface of SnTe.

\section*{Acknowledgement}
We acknowledge Roald Hoffmann for useful discussions.
Y.S., M.W. and J.F. are supported by China National Innovation Program, and F.Z. is supported by UT Dallas research enhancement funds and DARPA Grant No. SPAWAR N66001-11-1-4110.

\bibliography{SnTe}

\end{document}